\documentclass[reprint,aip,cha,floatfix,nofootinbib]{revtex4-1}

\usepackage{amsmath,amssymb}
\usepackage{soul}
\usepackage{cancel}
\usepackage{listings}
\usepackage{relsize}
\usepackage{graphicx}
\usepackage{xcolor}
\usepackage{hyperref}

\DeclareFontFamily{U}{mathx}{}
\DeclareFontShape{U}{mathx}{m}{n}{<-> mathx10}{}
\DeclareSymbolFont{mathx}{U}{mathx}{m}{n}
\DeclareMathAccent{\widehat}{0}{mathx}{"70}
\DeclareMathAccent{\widecheck}{0}{mathx}{"71}

\newcommand{\x}{\mathbf{x}}

\renewcommand{\selectlanguage}[1]{}
\newcommand{\pa}[1]{{#1}}

\newcommand{\ti}[1]{{#1}}

\begin{document}

\title{Modelling of the dewetting of ultra-thin liquid films on chemically patterned substrates: linear spectrum and deposition patterns}

\author{Tilman Richter}
\affiliation{
Helmholtz Institute Erlangen-N\"urnberg for Renewable Energy,
Forschungszentrum J\"ulich, Cauerstr.~1, 91058 Erlangen, Germany
}%
\author{Paolo Malgaretti}
\affiliation{
Helmholtz Institute Erlangen-N\"urnberg for Renewable Energy,
Forschungszentrum J\"ulich, Cauerstr.~1, 91058 Erlangen, Germany
}
\email{p.malgaretti@fz-juelich.de}
\author{Jens Harting}
\affiliation{
Helmholtz Institute Erlangen-N\"urnberg for Renewable Energy,
Forschungszentrum J\"ulich, Cauerstr.~1, 91058 Erlangen, Germany
}
\email{j.harting@fz-juelich.de}
\affiliation{Department of Chemical and Biological Engineering and Department of Physics, Friedrich-Alexander-Universität Erlangen-Nürnberg, Cauerstr.~1, 91058 Erlangen}


\begin{abstract}
Liquid films of nanometric thickness are prone to spinodal dewetting driven by disjoining pressure, meaning that a non-wetting liquid film of homogeneous thickness in the range of tens of nanometers will spontaneously break into droplets. The surface energy of the underlying solid substrate heavily influences the dynamics and resulting droplet configurations. Here, we study the dewetting of thin liquid films on physically flat but chemically heterogeneous substrates using the thin film equation. We use linear stability analysis (LSA) to describe and predict the system's behavior until the film ruptures and compare it to numerical simulations. 
\pa{The good agreement between the numerical solutions and the LSA allows us to %\ti{demonstrating that the dynamics of the liquid film prior to rupture are well predicted by the LSA}. 
%Based on our calculations, we 
propose a method for measuring surface energy patterns %. \ti{We show that it is theoretically possible to calculate the chemical patterning of the underlying substrate 
from early time-step film height profiles with good precision.} Furthermore, we study the non-linear dynamics and the eventually formed droplet pattern by numerical simulations. \ti{This offers insights into the dependency of the resultant droplet arrays on shape, feature size, and magnitude of the chemical patterning of the underlying substrate.}
\end{abstract}

\maketitle

\section{Introduction}
The wetting behavior of liquid films is the subject of long-lasting intensive research interest~\cite{oron_long-scale_1997, craster_dynamics_2009, de_gennes_capillarity_2004, becker_complex_2003, thiele1998dewetting, peschka_signatures_2019, snoeijer2013moving,singhpof:2022}. It is of great interest for many biological and technical applications, such as the embryogenesis of zebrafish yolks~\cite{wallmeyer_collective_2018}, the stability of the lachrymal film on the human eye~\cite{sharma_mechanism_1985}, supported liquid phase catalysts~\cite{richter_chemically_2024, frosch_wetting_2023}, protective coatings~\cite{kaita2010experiments, kasischke2023pattern}, printable photovoltaics~\cite{beynon_allprinted_2023, qiu_situ_2024, MQRLLDBEH25}, the friction of car tires on wet roads~\cite{de_gennes_capillarity_2004}, or many other applications~\cite{craster_dynamics_2009, oron_long-scale_1997, de_gennes_capillarity_2004, bonn2001wetting}.

In all the above-mentioned cases, when a thin film is deposited on a homogeneous solid substrate, it may dewet and form droplets depending on the solid-liquid, \pa{gas-liquid}, and \pa{solid-gas} surface energies, which also determine the contact angle of the eventual droplet~\cite{oron_long-scale_1997, peschka_signatures_2019, fetzer_slippage_2007, fetzer_thermal_2007, becker_complex_2003, vrij_possible_1966, doi_soft_2013, israelachvili_intermolecular_nodate, PSH22}. 
This dewetting phenomenon occurs \ti{because of} the effective (van der Waals) interactions between the solid-liquid and liquid-gas interfaces, which lead to "spinodal dewetting"%. \ti{The effect of spinodal dewetting} has been studied quite intensively in the past
~\cite{vrij_possible_1966, fetzer_thermal_2007, peschka_signatures_2019, oron_long-scale_1997}. For initially homogeneous liquid films of thickness below $100 \text{nm}$ the attractive van der Waals interaction between the gas and the solid results in an energetic potential which, in the case of partial wetting, can be large enough to nucleate holes in the film \cite{peschka_signatures_2019, de_gennes_capillarity_2004, israelachvili_intermolecular_nodate}. The resulting pressure contribution is called the disjoining pressure and depends on the strength of the three \pa{above mentioned} surface energies. 

This scenario becomes more involved when the surface of the solid substrate is not homogeneous. In fact, a non-constant solid-liquid interfacial energy or surface tension heavily influences the dynamics of the rupture process and the resulting droplet pattern. \pa{By means of different techniques, for example plasma etching~\cite{hinduja_scanning_2022}, it is possible to chemically pattern the solid substrate and hence to induce} %This may be used to craft 
patterns of dewetted liquids on solid substrates by stripes of various sizes of more and less wettable areas on the substrate~\cite{kargupta_templating_2001, kargupta_morphological_2002, zhang_how_2003,zhang_patterning_2003, luo_ordered_2004}. \pa{These techniques are used in lithographic printing\cite{lenz_competitive_2007} and/or in combination with switchable substrates which allow for the creation of chemical patterning on flat substrates on demand~\cite{umlandt_light-triggered_2022, ichimura_light-driven_2000, zitz_controlling_2023, grawitter_steering_2021, xin_reversibly_2010}}. %\ti{Also, for lithographic printing, the behaviour of thin liquid films on chemically patterned substrates is of high relevance \cite{lenz_competitive_2007}.} 
%\ti{Chemical patternings of substrates can occur naturally or may be created deliberately. For example, one can create a physically homogenous but chemically heterogeneous substrate by first plasma etching the desired pattern and then coating a thin, even layer on top, as done by Ref.~\cite{hinduja_scanning_2022}. Also, so-called switchable substrates allow for the creation of chemical patterning on flat substrates on demand~\cite{umlandt_light-triggered_2022, ichimura_light-driven_2000, zitz_controlling_2023, grawitter_steering_2021, xin_reversibly_2010}.} 

One \ti{commonly} employed tool to study wetting, especially spinodal dewetting, is the thin film equation (TFE), which allows for effective modeling of spinodal dewetting by including a so-called disjoining pressure, that can appropriately capture the physical mechanism leading to rupture of ultra-thin liquid films~\cite{craster_dynamics_2009, oron_long-scale_1997, peschka_signatures_2019}. A variety of different effects of patterned substrates on the spinodal dewetting of ultra-thin liquid films can be observed and explained by the TFE~\cite{kao_rupture_2006, thiele_modelling_2003}. However, to the best of our knowledge, the dynamics of the liquid film on a chemically patterned substrate leading to film rupture remains not fully understood.

By linear stability analysis (LSA) of the TFE, the spectrum of the film height profile of a thin liquid film on a physically flat and chemically homogeneous substrate undergoing spinodal rupture before film breakage can be well understood~\cite{zhang_linear_2024, zhang_molecular_2019, lam_computing_2019, gonzalez_inertial_2016, fetzer_thermal_2007, mecke_thermal_2005}. This has been validated against experiments~\cite{fetzer_thermal_2007} and molecular dynamics simulations~\cite{zhang_molecular_2019, zhang_linear_2024} demonstrating the validity of the linearised TFE to describe dewetting dynamics of thin liquid films. 

Furthermore, the influence of various functional forms of chemical patterning on solid substrates regarding the dewetting behavior of liquid films has been minimally explored and lacks comprehensive analysis. A notable attempt in this direction is Ref.~\cite{zitz_lattice_2021}, which investigates the spinodal dewetting of an ultra-thin liquid film on a solid substrate with heterogeneous local surface energy. The authors consider two types of substrate patterns: sinusoidal and step-function patterns. Their findings indicate that the resulting droplet pattern follows the substrate patterning. For sinusoidal patterns, droplets form only at the minima of the local contact angle, i.e., at the most wettable locations. This phenomenon is further illustrated in three dimensions in Ref.~\cite{zitz_controlling_2023}. For step-function patterns, droplets form centered on both hydrophobic and hydrophilic patches, with film rupture occurring at the discontinuities of the local contact angle $\theta$. As shown in Fig.~\ref{fig:dewetting_patterned}, new simulations have reproduced these results.
Additionally, Ref.~\cite{zitz_lattice_2021} notes exciting behavior in the spectrum of the film height (its Fourier transform) on patterned substrates, which was not further investigated. Ref.~\cite{zitz_lattice_2021} does not provide a more quantitative study of the dewetting behavior of liquid films on patterned substrates, nor do the authors present an explanation for the spectrum of the liquid film height profile upon dewetting on a chemically patterned substrate. In this article, we attempt to close this gap.

%The measurement of contact angles of sessile droplets is a common method for determining solid surface energies, known as contact angle goniometry~\cite{kwok_contact_1999, vuckovac_uncertainties_2019}. This concept can be advanced through scanning drop friction force microscopy (sDoFFI), which involves dragging a droplet on a glass capillary over a substrate and measuring the friction force to detect surface inhomogeneities smaller than $0.5 \text{mm}$~\cite{hinduja_scanning_2022}. Moreover, the dewetting dynamics of thin liquid polymer films have been proposed as a method for surface energy measurements \cite{choi_alternative_2003}, an idea that we expand on in this article.

This paper investigates the dynamics of a thin liquid film undergoing spinodal rupture on chemically patterned substrates. Our analysis encompasses both the linear regime and the nonlinear regime. To study the linear regime, we employ an LSA accounting for the patterning of the substrate. By numerical simulations, we verify the LSA and confirm that linear theory approximates the film dynamics before film rupture. To further study the nonlinear regime, we rely on numerical simulations. We perform parameter studies exploring the eventual steady-state droplet arrays that emerge after film rupture and how they depend on the shape of the underlying droplet pattern. Furthermore, we identify the underlying mechanism of film rupture by investigating the transition from one droplet per pattern wavelength on a sinusoidal pattern (Fig.~\ref{fig:dewetting_patterned}(a)) to two droplets per pattern wavelength on a step-function pattern (Fig.~\ref{fig:dewetting_patterned}(b)). Also, the effects of the pattern amplitude and wavelength are studied. 

Finally, we propose a novel method\pa{, similar to Ref~\cite{choi_alternative_2003},} for measuring surface energies by inverting the problem: deducing the chemical patterning of the substrate based solely on thin film dynamics. A comparison with numerical simulations shows that this method can accurately map surface energy heterogeneities of solid substrates. Our calculations offer a possible extension of the already explored wetting-based surface energy measurement techniques, contact angle goniometry~\cite{kwok_contact_1999, vuckovac_uncertainties_2019} and sDoFFI~\cite{hinduja_scanning_2022}. 

The remainder of this article unfolds as follows: In section~\ref{sec:Model}, we describe the employed TFE model. Section~\ref{sec:parameter_study} presents numerical results concerning the eventual steady state of droplets after film rupture, focusing on the number of droplets per pattern period of regular patterns and the rupture time.  In section~\ref{sec:lsa}, the LSA is derived and compared with numerical simulations. Finally, in section~\ref{sec:comparison}, we discuss the possibility of obtaining the substrate pattern from measurements of the thin film spectrum. 

\section{Model}
\label{sec:Model}

\begin{figure}
    \centering
    \includegraphics[width=\linewidth]{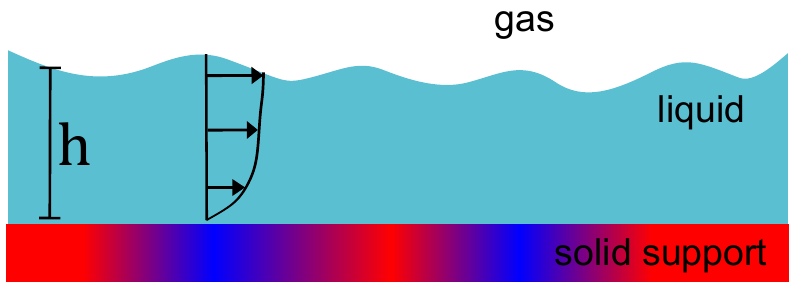}
    \caption{Sketch of the setup. The sketch shows the gas phase at the top. The liquid phase in blue is described by the film height $h$. The substrate below differs by the local surface energies expressed by the contact angle $\theta$ where blue-shaded areas have a low local surface tension, and red-shaded areas have a high local surface tension. }
    \label{fig:sketch}
\end{figure}

We employ the thin film equation (TFE) to study the dewetting of a thin liquid film resting on a solid substrate with heterogeneous wettability. The TFE is an effective conservation equation of the film height $h$, modeling its evolution in time, driven by the hydrodynamic pressure. It can be derived by approximation from the Navier-Stokes equation and is valid for liquid films having typical thicknesses smaller than any relevant horizontal length-scale. This assumption is called the lubrication approximation. Furthermore, the TFE is only valid in the Stokes regime where the Reynolds number is small. It reads~\cite{oron_long-scale_1997, craster_dynamics_2009}
\begin{align}
\label{eq:TFE}
\partial_t h = \nabla \cdot\left(\frac{M(h)h}{\mu}\nabla p\right).
\end{align}
In Eq.~\eqref{eq:TFE}, $\mu$ denotes the dynamic viscosity. The hydrodynamic pressure in the liquid film $p$ is assumed to be homogeneous along the transverse direction. It is given as a function of film height $h$, the sum of the Laplace pressure and the disjoining pressure. The Laplace pressure is driven by surface tension $\gamma$ and the curvature of the liquid interface and is given by $-\gamma \nabla^2 h$. The Laplace Beltrami operator $\nabla^2 h$ approximates the curvature of the film height profile valid in the lubrication approximation. The disjoining pressure is responsible for the spinodal dewetting. It stems from the van der Waals interactions between the surrounding gas atmosphere and the solid substrate and is given by the wetting potential $\varphi$. This wetting potential can be calculated by integrating the interaction potentials between molecules of the surrounding gas atmosphere and the solid substrate over the half-spaces of the atmosphere and the substrate. Employing a classical Lennard-Jones $(6,12)$ potential one derives the wetting potential to be~\cite{de_gennes_capillarity_2004, israelachvili_intermolecular_nodate, thiele_gradient_2016}
\begin{align}
\label{eq:wetting_potential}
   \varphi(h) &= \frac{1}{n-m}\left( n \left( \frac{h^*}{h}\right)^m - m \left( \frac{h^*}{h}\right)^n \right),
\end{align}
where $n=8$, and $m=2$ are two integers and $h^*$ is the precursor length. This functional form of the wetting potential gives rise to a minimal film thickness of $h^*$ \ti{by incorporating a short ranged repulsive force, closing a gap highlighted in the outlook of Ref.~\cite{lenz_competitive_2007}. This so-called precursor film }shall be considered as a dewetted film, i.e., whenever $h=h^*$ this has to be interpreted as the solid substrate being exposed to the surrounding gas atmosphere. The main purpose of this layer is to avoid division by zero when the liquid film dewets, even though there is evidence that, indeed, also in a real system, a layer of molecular thickness may be formed on dry substrates preceding liquid films or droplets~\cite{popescu_precursor_2012}. For charged interfaces, one would obtain instead $\varphi\propto \exp(-\kappa h)$ where $\kappa$ is the Debye length~\cite{israelachvili_intermolecular_nodate}. Including the surface energies, the disjoining pressure then reads $-\gamma(1-\cos \theta) \varphi'(h)$. Here, $\theta$ is the local contact angle of the liquid film given by Young's law \cite{doi_soft_2013}
\begin{align}
\label{eq:youngs_law}
    \cos \theta = \frac{\gamma_{sg}-\gamma_{sl}}{\gamma},
\end{align}
where $\gamma$, $\gamma_{sg}$, and $\gamma_{sl}$ are \ti{the three surface energies: gas-liquid, solid-gas, and solid-liquid}. To model the non-constant wettability of the solid substrate, we assume $\gamma_{sg}$ and $\gamma_{sl}$ to be dependent on space and thus also $\theta$ being a non-constant function of space. \ti{Within our model, the contact angle is defined as Young's angle obtained by Eq.~\eqref{eq:youngs_law}. Thus, our model always assumes Young's angle at equilibrium up to numerical precision and does not reproduce the effect of contact angle hysteresis observed in experiments~\cite{butt_physics_2003, eral_contact_2013}. Furthermore, we assume, for simplicity, perfectly smooth substrates. The roughness of the substrate can be incorporated into Eq.~\eqref{eq:youngs_law} using Wenzel's relation~\cite{de_gennes_capillarity_2004}. Therefore, strictly speaking, the employed model captures heterogeneities of the apparent contact angle independent of whether they stem from variations in solid-liquid surface energy or local roughness of the substrate.} $\varphi'(h)$ being the derivative of the wetting potential $\varphi$, the hydrodynamic pressure reads 
\begin{align}
\label{eq:pressure}
    p(h)= - \gamma \nabla^2 h -\gamma(1-\cos \theta) 
    \varphi'(h).
\end{align}

Finally, to fully explain Eq.~\eqref{eq:TFE}, we introduce the mobility $M(h)$ given by 
\begin{align}
    \label{eq:mobility}
    M(h)=\frac{2h^2+6bh+3b^2}{6},
\end{align}
where $b$ is the so-called slip length. Eq.~\eqref{eq:mobility} assumes that the velocity profile 
$u(\mathbf{x},z)$ in the liquid film to be of a parabolic shape. Such an assumption relies on the fact that the pressure inhomogeneity along the film induces the fluid flow and that we assumed the pressure to be homogeneous along the transverse direction. On top of this, the following boundary conditions are imposed: a) slip boundary at the solid substrate $u(\x, -b)=0$, i.e., the liquid is allowed to move over the solid substrate as observed experimentally~\cite{peschka_signatures_2019, fetzer_slippage_2007, rauscher_spinodal_2008, zhao_slip_2002} (one can, of course, choose $b=0$ to restore the no-slip boundary condition. However, small non-zero values of slip $b$ do not substantially influence the results but increase the numerical stability of the model),  and b) non-viscous atmosphere $\partial_z u(\x, h)=0$, i.e., no dissipation at the liquid-gas interface. 

In this manuscript, all numerical results are obtained by solving Eq.~\eqref{eq:TFE} by a newly developed 
lattice Boltzmann method for thin-liquid-film hydrodynamics presented \ti{and validated} in Refs.~\cite{zitz_lattice_2019, zitz_lattice_2021, zitz_swalbejl_2022, zitz_controlling_2023}. 

\section{Dependence of the Emerging Droplet Pattern and Rupture Time on the Substrate Patterning}
\label{sec:parameter_study}

\begin{figure*}
    \centering
   \includegraphics[width=\linewidth]{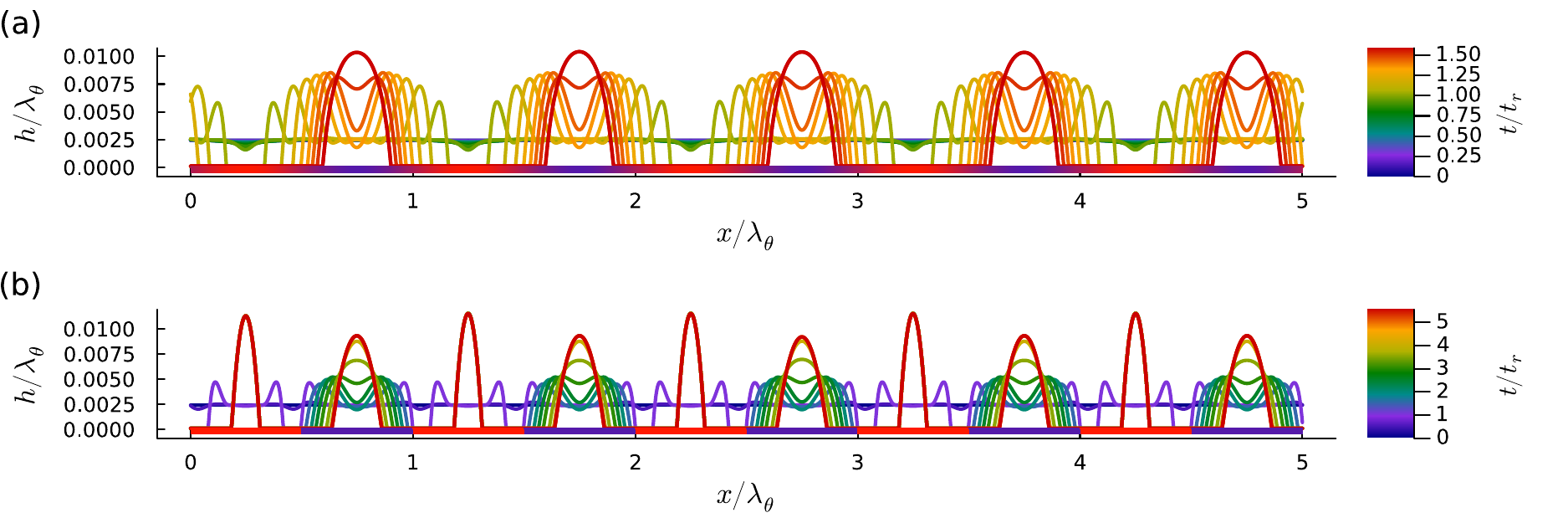}
    \caption{Simulations of the dewetting on a patterned substrate. All lengths are reported in units of the pattern wavelength $\lambda_{\theta}$. Lines of different colors represent the liquid film height $h$ at different time steps. The substrate below is plotted as a colored bar where blue indicates a low local contact angle $\theta$ and red a high local contact angle. All parameters, including the mean contact angle $\theta_0$, its amplitude $\delta \theta$, and wavelength $\lambda_\theta$, are the same in both simulations shown; they only differ by the functional form of the pattern. Here, we have chosen the pattern wavelength to be twice the spinodal wavelength $\lambda_\theta=2/q_{\max,plain}$ (see Eq.~\eqref{eq:q_max}). (a) simulation of the dewetting on a sine-wave pattern (b) simulation of the dewetting on a step-function pattern }
    \label{fig:dewetting_patterned}
\end{figure*}

In this section, we investigate the breakage of liquid films on chemically patterned substrates and the emerging droplet patterns by numerical simulations. We systematically study the influence of the shape of the substrate pattern on the number of droplets formed after film rupture and the time it takes the system to evolve from the initially flat film to the formation of the first hole (the so-called rupture time $t_r$). 

To illustrate the breakage of a liquid film on a patterned substrate, we show two exemplary simulations in Fig.~\ref{fig:dewetting_patterned}, solving the model numerically on differently patterned substrates. Similar simulations have been performed in Ref.~\cite{zitz_lattice_2021}. The initial condition is, as for all simulations presented in this paper, a constant film height with small random perturbations $h(x,t=0)=h_0 + \epsilon\mathcal{N}(x)$. In Fig.~\ref{fig:dewetting_patterned}, we use a pattern wavelength $\lambda_\theta=2/q_{\max, \text{plain}}$, where $q_{\max, \text{plain}}$ is the fastest growing mode of a thin liquid film on a homogeneous substrate (See section~\ref{sec:lsa}, Eq.~\eqref{eq:q_max}). 

In Fig.~\ref{fig:dewetting_patterned}(a), we use a sine-wave pattern $\theta=\theta_0 + \delta \theta \sin(2 \pi x/ \lambda_\theta)$, where $\theta_0=\pi/9=20^\circ$ and $\delta\theta=\pi/19=10^\circ$. We observe the film breaking at the maxima of the pattern $\theta$ where the substrate is least wettable. From there on, wetting rims advance, forming a single droplet when meeting. This leads to the formation of one droplet per pattern period at the position of the smallest local contact angle $\theta$. 

In Fig.~\ref{fig:dewetting_patterned}(b), a step-function pattern $\theta= \theta_0+\delta\theta\lbrace2\Theta[\text{mod}(x-\lambda_\theta/2, \lambda_\theta)] -1\rbrace$ is employed. Here, $\Theta$ denotes the Heavyside function. The film breaks at the step of the pattern, sending one wetting rim traveling into the red region of high local contact angle and one into the blue region of low local contact angle. Two droplets are formed, one on the more hydrophobic part of the substrate and one on the more hydrophilic part. 

The difference in the number of formed droplets between the sine wave pattern in panel (a) and the step function in panel (b) can be explained as follows: By the evolution equation Eq.~\eqref{eq:TFE}, the only mechanism that can lead to film rupture driven by local differences in the substrate surface energy, is by gradients in the pressure $p$. When the film height is constant in the initial condition, those gradients in the pressure stem from gradients in the local contact angle $\theta$. In the case of the sine wave, the liquid film breaks at the maxima of the pattern at a \emph{spot-defect} of the substrate's surface energy. For the step function, on the other hand, at the maximum of the pattern, the pressure is constant almost everywhere as the contact angle is constant almost everywhere. The only location where one has a gradient in the contact angle and, thus, in pressure is at the step of the chemical surface pattern. The liquid film breaks at this \emph{step-defect} of the substrate's surface energy. 
%\jh{spot-step-defect?}

The above observation that a thin film is forming different amounts of droplets on sinusoidal substrate patterns or step function surface patterns driven by spot- or step-defects of the substrate's surface energy raises some questions: What happens when the step of the step function is smoothened, as in an experiment a pattern never will be arbitrarily sharp? When morphing from a step-function pattern to a sinusoidal pattern, at which point do we observe a transition between the emergence of two droplets per pattern period and one droplet per pattern period? How do the results depend on the pattern wavelength $\lambda_\theta$ and the pattern amplitude $\delta\theta$? In the following two subsections, we will answer those questions by parameter studies in numerical simulations. 

\subsection{Transition from Sine Wave to Step Function Patterns}

We study the transition between the different amounts of droplets that might emerge after the rupture of a liquid film on the differently patterned substrates shown in Fig.~\ref{fig:dewetting_patterned}. We are especially interested in when the liquid film breaks on spot-defects as it is the case for the sine wave pattern with a wavelength of $\lambda_\theta=2 q_{\max,plain}$ reported in Fig.~\ref{fig:dewetting_patterned}(a) and when it breaks at the step-defects as seen in Fig.~\ref{fig:dewetting_patterned}(b). In agreement with Ref.~\cite{zitz_lattice_2021}, we observe, depending on the pattern used, one or two droplets per substrate-pattern wavelength. Therefore, we expect a transition between the emergence of one or two droplets when transitioning from a perfect step-function pattern to one resembling a sine wave more closely.

\begin{figure}[h!]
    \centering
    % \begin{minipage}{\linewidth}
    % \subnumber{(a)}
    %     \includegraphics[width=\linewidth]{patterned_smoothed_theta.pdf}
    % % \label{fig:patterned_smoothed_theta}
    % \end{minipage}
    % \hfill
   \begin{minipage}{\linewidth}
   % \subnumber{(b)}
    \includegraphics[width=\linewidth]{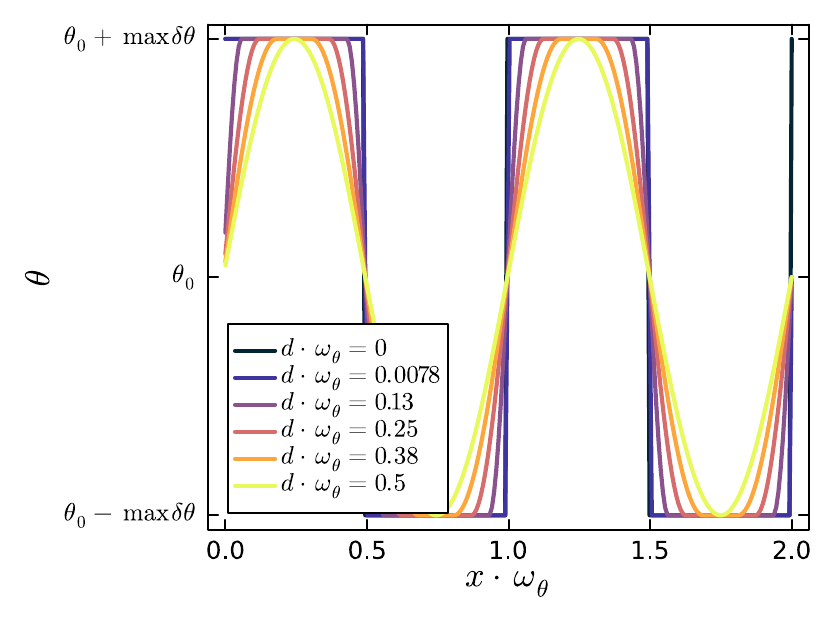}
    % \label{fig:patterned_smoothed_theta_d}
   \end{minipage}
   % \caption{(a) The substrate pattern contact angle $\theta$ given by Eq.~\eqref{eq:patterned_smoothed_pattern} is a step function with steps smoothened out by a sine-wave, with average contact angle $\theta_0$, amplitude $\max \delta \theta$ and the smoothing width $d$. (b) The substrate contact angle $\theta$ as in panel (a) upon varying the smoothing width $d$ with respect to the substrate wavenumber $\omega_\theta$, showing a transition from a step function at $d=0$ to a sine-wave at $d=1/\omega_\theta$.}
   \caption{\ti{The substrate pattern contact angle $\theta$ given by Eq.~\eqref{eq:patterned_smoothed_pattern} is a step function with steps smoothened out by a sine-wave, with average contact angle $\theta_0$, amplitude $\max \delta \theta$ and the smoothing width $d$. Lengths are reported in terms of the substrate wavenumber $\omega_\theta$. Upon varying the smoothing width $d$, we see a transition from a step function at $d=0$ to a sine-wave at $d=1/\omega_\theta$.}}
   \label{fig:patterned_smoothed_pattern}
\end{figure}

To study this expected transition quantitatively, we construct a substrate patterning as shown in Fig.~\ref{fig:patterned_smoothed_pattern}, i.e., we use a local contact angle given by 
\begin{widetext}
\begin{align}
\label{eq:patterned_smoothed_pattern}
	\theta(x)=\begin{cases}
		\theta_0 + \sin(x\pi /d)\cdot \max \delta \theta, & k\lambda_\theta\leq x<k\lambda_\theta+d/2\\
		\theta_0 + \max \delta \theta, &k\lambda_\theta + \ d/2\leq x < k\lambda_\theta + \lambda_\theta/2 - d/2\\
		  \theta_0 + \sin(\pi(x-\frac{\lambda_\theta-d}{2})/d+\frac{\pi}{2})\max\delta \theta, & k\lambda_\theta + \lambda_\theta/2 - d/2 \leq x < k\lambda_\theta +  \lambda_\theta/2 + d/2 \\
		\theta_0 - \max \delta \theta, & k\lambda_\theta + \lambda_\theta/2 + d/2 \leq x <k\lambda_\theta + \lambda_\theta - d/2\\
		\theta_0 + \sin(\pi(x-\lambda_\theta+d/2)/2)\max \delta \theta, &k\lambda_\theta +  \lambda_\theta - d/2 \leq x < k\lambda_\theta + \lambda_\theta, 
	\end{cases}
\end{align}
\end{widetext}
where $k \in \mathbb{N}$ is a number. With this definition, we smoothen out the step function pattern by inserting a quarter sine-wave with width $d$ instead of the discontinuity. We denote $d$ the \emph{smoothing width} of the pattern. In the limit $2d=\lambda_\theta$, we obtain a sine-wave and for $d=0$ a step-function.

\begin{figure*}
    \centering
    \includegraphics[width=\linewidth]{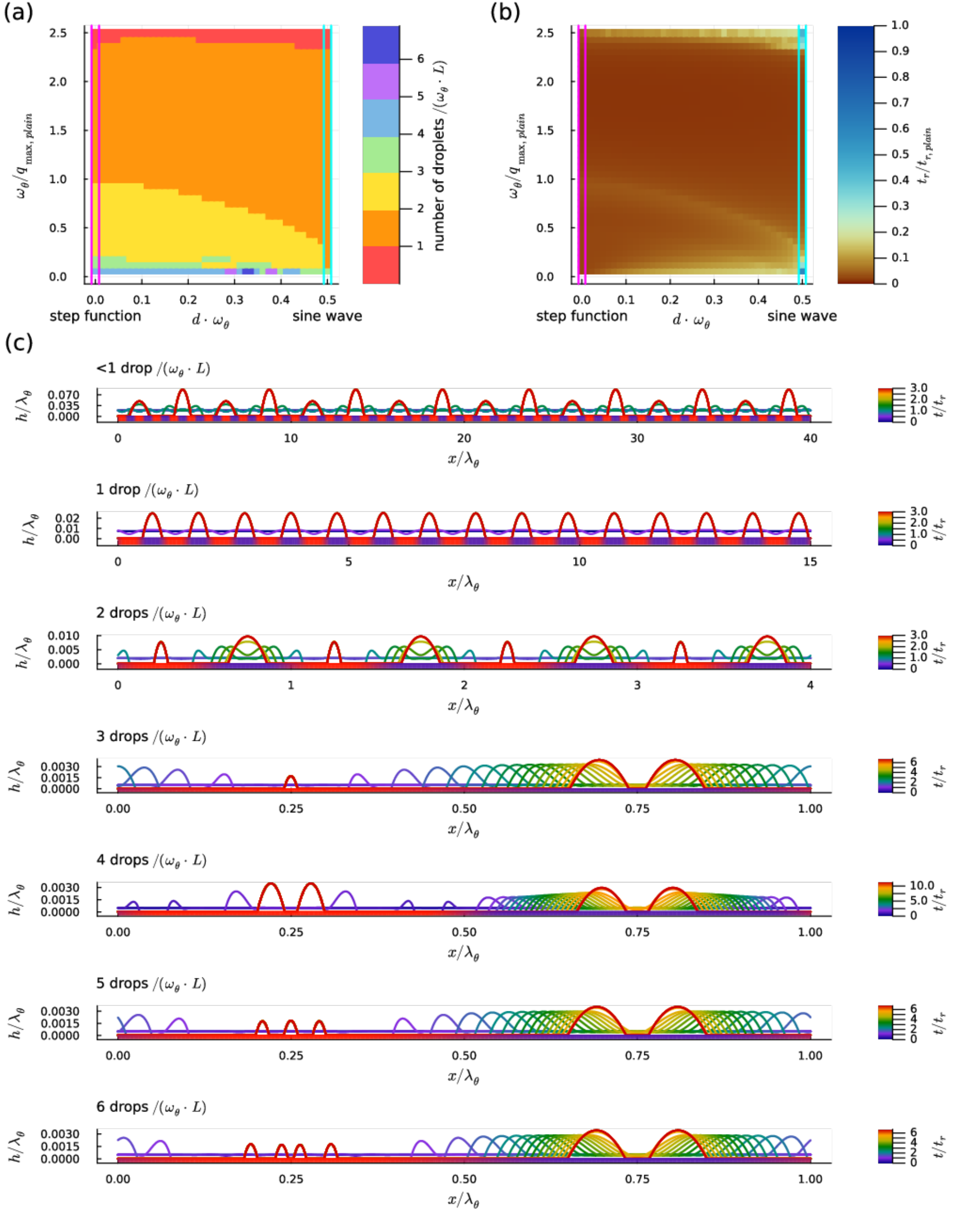}
    \caption{Varying the smoothing width $d$ and the pattern wavenumber $\omega_\theta$ normalized by the fastest growing mode of dewetting on the plain substrate $q_{\max, plain}$. Data reported between cyan and magenta lines correspond to the same parameters in Fig.~\ref{fig:sin_step_params}. (a) Number of droplets formed after film rupture. (b) Time $t_r$ for film rupture to occur, normalized by the rupture time on a homogeneous substrate $t_{r,plain}$. (c) Snapshots of selected exemplary simulations showing the emergence of different amounts of droplets after film rupture. }
    \label{fig:patterned_parameter_study}
\end{figure*}

\newcommand{\omegalimit}{{2.5}}
\newcommand{\omegathresh}{}

We perform a parameter study over the smoothing width $d\in [0, \lambda_\theta]$ and the wavenumber of the substrate pattern $1/\lambda_\theta=:\omega_\theta \in (0, \omegalimit q_{\max,plain}]$. The results of this study, including $41\cdot 40=1640$ simulations in total, are visualized in Fig.~\ref{fig:patterned_parameter_study}. Here, inset (a) reports the number of droplets per period of the substrate pattern.

At roughly $\omega_\theta\gtrsim \omegalimit q_{\max,plain}$, we observe that the number of emerging droplets per substrate period is less than one, meaning that a single droplet or hole covers more than a single period of the patterning. This phenomenon for large substrate pattern wavenumbers has also been reported by Ref.~\cite{kargupta_dewetting_2002}. Interestingly, the number of droplets observed after film rupture attains a metastable state containing a number of droplets that fit the dominant wavenumber $q_{\max,plain}$ of a substrate with a constant local contact angle equal to the average contact angle $\theta_0$ of the patterned substrate.

Regarding the emerging arrays of droplets, one might conclude that the substrate can be treated based on its average properties. This is not entirely true for the pattern wavelengths studied here, as we will see later when discussing the rupture time. Still, it is a good approximation when one is only interested in the droplet patterns emerging from spinodal dewetting.

Fig.~\ref{fig:patterned_parameter_study}(c) shows snapshots of an exemplary simulation from the regime $\omega_\theta\gtrsim \omegalimit q_{\max,plain}$ in the first row (labeled $<1 \text{drop} / (\omega_\theta \cdot L)$). 
For $\omegalimit q_{\max,plain} \gtrsim \omega_\theta>\omegathresh q_{\max, plain}$ we observe a single droplet per period of the substrate pattern after film rupture, independent of the smoothing width $d$. In this regime, the local maxima of the contact angle $\theta$, i.e., the places of low wettability, act as spot-defects of the substrate, provoking the nucleation of a single hole per defect and thus per period of the substrate pattern.
For $\omega_\theta<\omegathresh q_{\max,plain}$ we then observe the expected transition between one droplet per period for a sine-wave pattern ($d\cdot \omega_\theta=d/\lambda_\theta=1/2$) and two droplets for a step-function pattern ($d \omega_\theta=0$). The transition value of $d$ is the point where the smoothed-out step no longer acts on the liquid film as a step-defect, which creates a hole per step in the liquid film. Instead, the maximum of the local contact angle begins to act as a single spot-defect, at which the film breaks. Fig.~\ref{fig:patterned_parameter_study}(c) shows snapshots of example simulations illustrating this principle in the second and third rows (labeled $1~\text{drop} / (\omega_\theta \cdot L)$ or $2~\text{drop} / (\omega_\theta \cdot L)$, respectively).
Upon decreasing $\omega_\theta$ further, we observe three or more droplets per pattern period for the smallest pattern-wavenumbers considered in our simulation set. This is best understood by looking at snapshots of corresponding simulations as shown in the last four rows of Fig.~\ref{fig:patterned_parameter_study}(c) (labeled $3-6~\text{drops} / (\omega_\theta \cdot L)$). The film breaks at the location of the highest hydrophobicity (red-colored substrate), leaving behind one or even more droplets.
The number of droplets formed at the location of highest hydrophobicity is, to a certain degree, dependent on the random seeding of the simulation but can partly be explained by the specific pattern at hand.

First, we realize that for smaller wavelengths of the substrate patterning, a single droplet is left on the hydrophobic part of the substrate. Only upon increasing the substrate-pattern wavelength is there room for more droplets. 
To understand the number of emerging droplets after film rupture, we describe the formation of droplets on the hydrophobic part of the substrate for different values of the smoothing width $d$ when the wavelength of the pattern is large enough to allow for multiple droplets on the hydrophobic part: 

For small $d$, i.e., step-function-like patterns, the film breaks first at the step of the substrate pattern. From the holes created in the process, two wetting rims travel inwards towards the location of the highest hydrophobicity, and two wetting rims travel towards the places of the highest hydrophilicity. When the two rims meet on the hydrophobic part of the substrate, the film breaks in between them, forming two droplets. This process has been intensively described and studied in Ref.~\cite{diez_breakup_2007}. The same happens on the hydrophilic part of the substrate: once the two wetting rims meet, the liquid film in between breaks. 

For higher $d$, the step in the substrate pattern is smoothened out such that the film breaks on the hydrophobic area, forming 2, 3, 4, or even 5 holes, creating 1, 2, 3, or 4 droplets. The larger the smoothing width $d$, the smaller the area of high constant high local contact angle $\theta$, and thus, the number of formed droplets on the hydrophobic part of the substrate is largest for intermediate smoothing widths $d$ and decreases again for very large $d$.

Having described the formation of multiple droplets on the hydrophobic part of the substrate, all that is left to understand is to obtain the total number of droplets per substrate pattern period, which is the number of droplets on the hydrophilic part of the substrate. Two droplets are formed for large substrate pattern wavelengths on the hydrophilic part. This can be understood as already described. When the film breaks on the step of the substrate pattern or the hydrophobic part of the substrate, two wetting rims travel toward the hydrophilic part of the substrate. Their movement is sped up as they follow the wettability gradient of the substrate. When the two rims meet, the film in between breaks and creates two more droplets. 

Next, we describe the rupture time measured by our simulation set presented in Fig.~\ref{fig:patterned_parameter_study}(b). In general, it can be said that rupture is significantly sped up by chemical patterning.
%, as already seen in the last chapter, where we have seen how rupture time can be obtained to a certain approximation from the LSA. 
As for the dependency on the control parameters smoothing width $d$ and pattern wavenumber $\omega_\theta$ many of the features of the diagram showing the number of droplets after film rupture (Fig.~\ref{fig:patterned_parameter_study}(a)) can be seen again for the rupture time.

For $\omega_\theta\approx \omegalimit q_{\max,plain}$, where less than one droplet is formed per substrate pattern period, the film takes about $t_r\approx 0.2 t_{r,plain}$ to rupture. Here, $t_{r,plain}$ denotes the rupture time of a thin film of the same parameters on a homogeneous contact angle $\theta_0$ substrate. If, additionally, the smoothing width is $d\cdot \omega_\theta=1/2$, i.e., for a perfect sine-wave substrate pattern, the rupture time $t_{r,plain}$ of the plain pattern is almost recovered.

In the region of $\omega_\theta\lesssim \omegalimit q_{\max, plain}$ where one droplet is formed per substrate pattern period, we observe much smaller rupture times $t_r<0.1 t_{r,plain}$. Here, the rupture is sped up by the spot-like defects of the substrate. 

A local maximum of the rupture time of $t_r\approx 0.175 t_{r,plain}$ is observed at the transition between the one and the two droplet regions in the diagram. This is where the system transitions from rupture driven by spot- or step-defects of the substrate. Consequently, the rupture is not that strongly sped up by either one of those defects. 

Going further down into the two-droplet region in the diagram, the rupture time drops again to $t_r<0.1t_{r,plain}$, as the rupture is now driven by pronounced step-like defects of the substrate. 
For the smallest values of the substrate pattern wavelength $\omega_\theta$ considered here and especially for larger values of the smoothing width $d$, the step in the substrate pattern is less pronounced, and as a consequence, the rupture time is raised again to $t_r>0.15 t_{r,plain}$. For the most extreme case of the smallest given value of $\omega_\theta$ and $d\cdot \omega_\theta=1/2$, i.e., a perfect sine-wave spanning the whole simulation box, the rupture time of the plain substrate $t_{r,plain}$ is almost recovered.

\ti{The results presented here are independent on the precursor height $h^*$ as shown in the supporting Figures S1, S2, S3.}

\subsection{Amplitude and Pattern-wavelength}

\begin{figure}
    \centering
    \includegraphics[width=\linewidth]{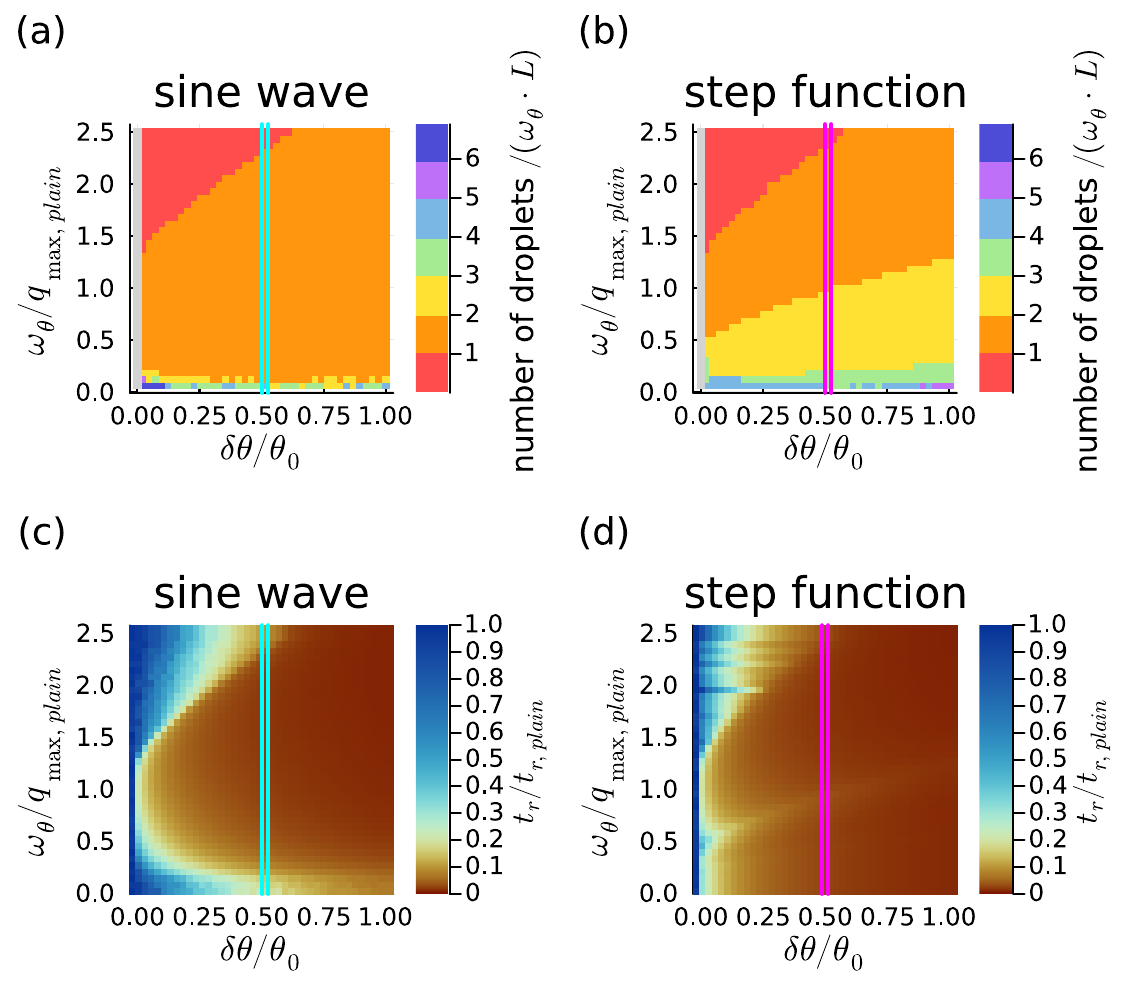}
    \caption{Varying the pattern wavenumber $\omega_\theta$ and the pattern amplitude $\delta\theta$ for sine wave patterns (left) and step function patterns (right). Data reported in between cyan and magenta lines correspond to the same parameters in Figs.~\ref{fig:patterned_parameter_study}(a) and (b).  (a-b) The number of droplets formed after film rupture for (a) a sine-wave pattern and (b) a step-function pattern. \ti{For $\delta \theta = 0$ the substrate is homogenous, thus, the number of droplets can't be reported in units of the substrate pattern period. This situation is reported in grey.} (c-d) Time $t_r$ for film rupture to occur, normalized by the rupture time on a homogeneous substrate $t_{r,plain}$ for (c) a sine-wave pattern and (d) a step-function pattern.}
    \label{fig:sin_step_params}
\end{figure}

In this subsection, we study the dependency of the number of formed droplets after dewetting, as well as the time until film rupture on the amplitude $\delta \theta$ and the wavelength $\lambda_\theta$ or wavenumber $\omega_\theta=1/\lambda_\theta$, respectively, of the pattern used. To do so, we perform a parameter study varying the pattern-wavenumber $\omega_\theta\in (0,\omegalimit q_{\max,plain}]$, and the amplitude $\delta \theta\in[0, \theta_0]$, for a sine-wave pattern and a step-function pattern running $40\cdot 40=1600$ simulation for each kind of pattern. The results of this study are reported in Fig.~\ref{fig:sin_step_params}. 
The simulations performed in the previous subsection and the ones performed here overlap when the pattern amplitude is chosen $\delta \theta=0.5\theta_0$ and the smoothing width $d/\omega_\theta=1/2$ for the sine wave and $d=0$ for the step function pattern. This is indicated by colored lines in Figs.~\ref{fig:patterned_parameter_study} and \ref{fig:sin_step_params}. 

Fig.~\ref{fig:sin_step_params}(a) reports the number of droplets per period of the employed substrate pattern observed immediately after film-rupture. In the top left corner of the diagrams, i.e., for small pattern wavelength $\lambda_\theta=1/\omega_\theta$ and small pattern amplitude $\delta \theta$, we consistently produce less than a single droplet per pattern period. Thus, the emerged droplets span multiple periods of the pattern. Upon raising the pattern amplitude $\delta \theta$ or the pattern wavelength $\lambda_\theta=1/\omega_\theta$, the spinodal rupture starts following the substrate pattern. A single droplet is formed at the minimum local contact angle. The liquid film perceives the point of highest hydrophobicity as a spot-defect and breaks at this point, nucleating a hole. For the sine-wave pattern, more than one droplet is observed only for the most extended pattern wavelengths simulated. For the step function pattern, on the other hand, when the pattern is large enough, the discontinuity of the pattern acts as a step defect, creating a hole in the liquid film. Consequently, two droplets emerge, one on the more hydrophobic part and one on the more hydrophilic part. Again, more droplets can be observed for the largest pattern wavelengths simulated. 

The rupture time obtained by the parameter study is reported in Fig.~\ref{fig:sin_step_params}(b). For $\delta\theta=0$, the substrate is homogeneous, and thus, the rupture time of a film on a plain pattern is recovered $t_r=t_{r,plain}$. In the top left corner of the diagrams, for small patterns with small amplitudes, the rupture time decreases toward the transition line between patterns that produce less than one droplet and exactly one droplet. For the sine wave patterns, the rupture increases again when approaching the largest simulated pattern wave length $\lambda_\theta=1/\omega_\theta$ (bottom left corner). This does not happen for the step function pattern. Furthermore, we observe a local maximum of rupture time for the step function pattern at the transition between patterns that produce one or two droplets when the dynamics switch from breaking at a spot defect to breaking at a step defect.

\section{Linear Stability Analysis (LSA)}
\label{sec:lsa}

\begin{figure}
    \centering
     \includegraphics[width=\linewidth]{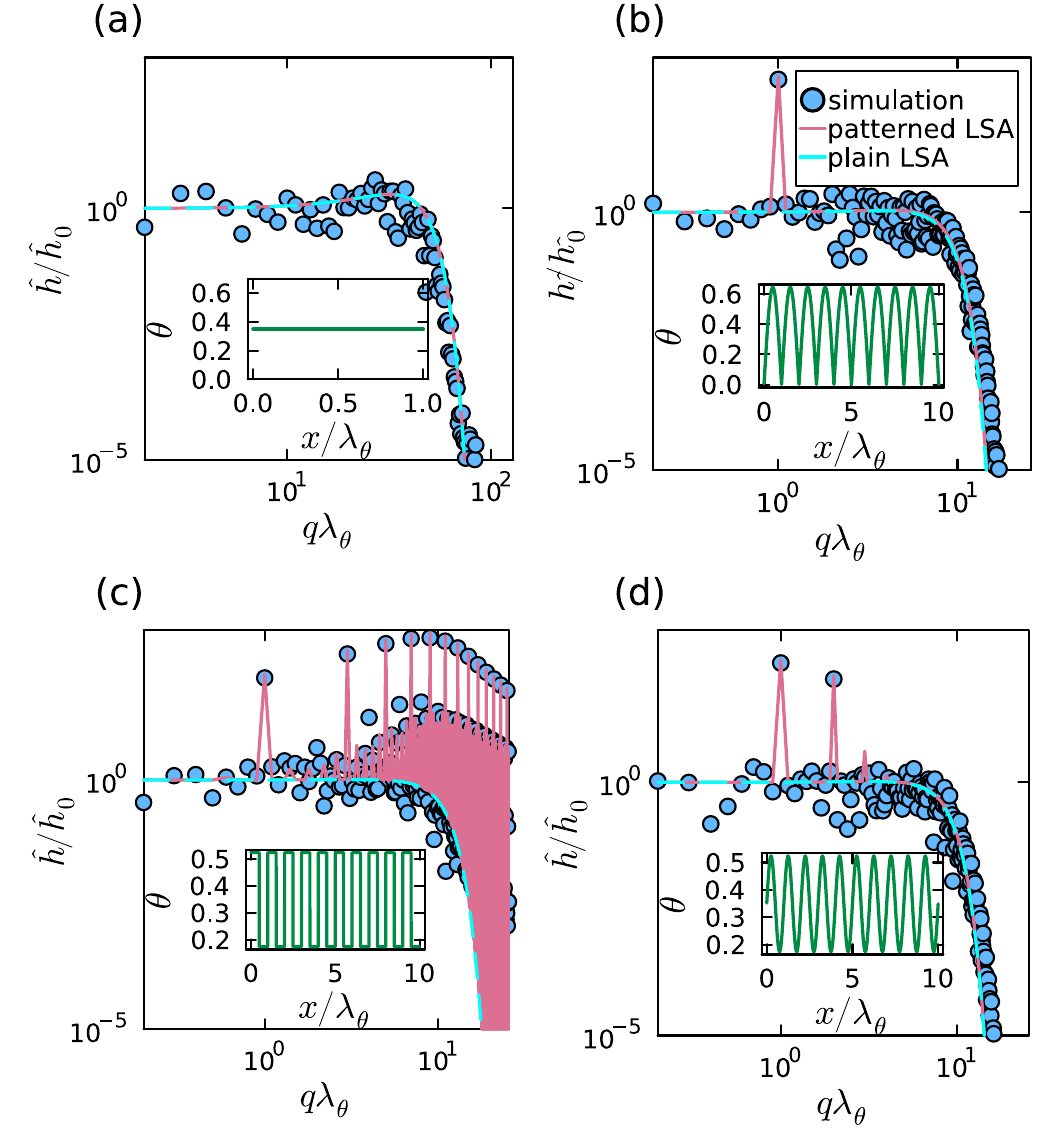}
	\caption{Early time-step simulation snapshots of dewetting on solid substrates with four different chemical patternings started from a homogeneous film-height with small perturbation $h(x,0)=h_0+\epsilon \mathcal{N}$ where $\mathcal{N}$ is a random variable. The spectrum (fast Fourier transform (FFT) of the data) of $h$ is plotted as dots, and the result of our calculation Eq.~\eqref{eq:solution_linear_patterned_0} as a solid violet line while the linear solution of the TFE without chemical patterning ($\delta \cos\theta=0$, Eq.~\eqref{eq:lsa_linear_solution}) is shown as a dashed cyan line. The horizontal axis shows the wavenumber $q=1/\lambda$ for $\lambda$ the corresponding wavelength multiplied by the wavelength of the employed pattern $\lambda_\theta$ and the vertical axis shows the relative excitement $\widehat{h}/\widehat{h_0}$ of the wavemode with wavenumber $q$, where  $\widehat{h}_0$ is the average value of the spectrum of $h$ at the initial condition. In real space, the local contact angle $\theta$ is shown as a green line in the insets. Panels (a)-(d) show different chemical patterns of the solid substrate.}
    % \pa{PM: Here the labeling is ok. Just "embed" the labels within the area of the plot (as much as possible, yet in a homogeneous manner among the panels}}
\label{fig:spec}
\end{figure}

After studying the rupture time and the emerging number of droplets after film rupture, we investigate the dynamics leading to film rupture in this section. 
To do so, we employ a linear stability analysis (LSA) to study the spectrum obtained from an ultra-thin liquid film dewetting on a chemically patterned substrate. As long as the deviation of the film homogeneity is small, we expect a good approximation of the spectrum of the film height $h$. 

Let us consider the TFE Eq.~\eqref{eq:TFE} and expand the height $h$ around a homogeneous value $h_0$ with perturbation $\delta h$:
\begin{align}
\label{eq:h_delta}
h= h_0 + \delta h
\end{align}
Similarly, we expand the wettability of the substrate $\cos \theta$ around the mean value $(\cos \theta)_0=\cos \theta_0 $ with a perturbation $ \delta(\cos \theta)$
\begin{align}
\label{eq:contact_angle_delta}
\cos \theta= (\cos \theta)_0+ \delta(\cos \theta).
\end{align}
Eq.~\eqref{eq:contact_angle_delta} is the main idea behind the calculation presented here. Treating the wettability in this way allows us to access the spectrum of a liquid film dewetting on a chemically patterned substrate.  We linearize Eq.~\eqref{eq:TFE} under the ansatz Eq.~\eqref{eq:h_delta}-\eqref{eq:contact_angle_delta} yielding

\begin{align}
\begin{split}
    \partial_t \delta h = \frac{\gamma M(h_0)h_0}{\mu} 
%\nonumber\\
\nabla\!\cdot\! \Huge[& -(1\!-\!(\cos \theta)_0)\varphi''(h_0)\nabla \delta h\! \\&+\! \varphi'(h_0)\nabla \delta (\cos \theta) \!-\! \nabla^3 \delta h \Huge].
\end{split}
\end{align}
The next step is to Fourier transform to $\widehat{\delta h}= \widehat{\delta h}(q)=\widehat{ h}(q), \widehat{\delta (\cos \theta)}= \widehat{\delta (\cos \theta)}(q)=\widehat{ \cos \theta}(q)$. Following the notation of Evans~\cite{evans2022partial}, we use the hat $\widehat{\quad}$ over the respective symbols as a notation for the Fourier transform
\begin{align}
\label{eq:fourier_transform}
    \widehat{f(x)}(q):=\int f(x) \exp(-i2\pi qx) dx,
\end{align}
and the inverse hat $\widecheck\quad$ for the inverse Fourier transform
\begin{align}
\label{eq:inverse_fouriere}
    \widecheck{g(q)}(x)=\int g(q) \exp(i2\pi qx) dq.
\end{align}
$q$ is the wavenumber given by $q=1/\lambda$ for the wavelength $\lambda$. 
This Fourier transform leads to  

\begin{align} 
\label{eq:patterned_fourier}
\begin{split}
    \partial_t \widehat{\delta h}= \frac{\gamma M(h_0)h_0}{\mu}
%\nonumber\\
\Huge[& q^2(1-(\cos\theta)_0)\varphi''(h_0)\widehat{\delta h}\\&- q^2 \varphi'(h_0)\widehat{\delta(\cos \theta)}- q^4 \widehat{\delta h} \Huge].
\end{split}
\end{align}
We can rewrite Eq.~\eqref{eq:patterned_fourier} as
\begin{align}
\label{eq:paterned_fourier_0}
\partial_t \widehat{\delta h}= \sigma(q) \widehat{\delta h }- q^2\frac{\gamma M(h_0)h_0}{\mu}\varphi'(h_0)\widehat{\delta(\cos\theta)},
\end{align}
where we have collected the dispersion relation or structure factor for the unperturbed case ($\delta\cos\theta=0$): 
\begin{align} 
\sigma(q) = \frac{\gamma M(h_0)h_0}{\mu}\left[ q^2(1-(\cos\theta)_0)\varphi''(h_0)- q^4  \right].
\end{align}
Equation~\eqref{eq:paterned_fourier_0} is an ODE and, as such, can be easily solved to obtain the solution
% \begin{align}
% \label{eq:solution_linear_patterned}
% \widehat{h}(q)=S_0 e^{\sigma(q) t}+ \frac{q^2 \frac{\gamma M(h_0)h_0}{\mu}\varphi'(h_0)\widehat{\cos\theta}}{\sigma(q)}.
% \end{align}
% The factor $S_0$ is obtained from the initial condition. It reads 
% \begin{align}
% S_0 = \widehat{h}(q,t=0)-\frac{q^2 \frac{\gamma M(h_0)h_0}{\mu}\varphi'(h_0)\widehat{\cos\theta}}{\sigma(q)}.
% \end{align}
% We can rewrite Eq.~\eqref{eq:solution_linear_patterned} as
\begin{align}
\label{eq:solution_linear_patterned_0}
&\widehat{h}(q,t)= \widehat{h}(q,t=0)e^{\sigma(q) t}\nonumber\\
&-\frac{(e^{\sigma(q)t}-1)}{\sigma(q)}q^2 \frac{\gamma M(h_0)h_0}{\mu}\varphi'(h_0)\widehat{\cos\theta}.
\end{align}
Choosing the perturbation of the local contact angle  $\delta \cos \theta=0$, we obtain the spectrum of a dewetting film on a homogeneous "plain" substrate \cite{oron_long-scale_1997}
\begin{align}
\label{eq:lsa_linear_solution}
\widehat{h}(q)= \widehat{h}(q,t=0)e^{\sigma(q) t}. 
\end{align}

Equation~\eqref{eq:solution_linear_patterned_0} illustrates the structure of the obtained spectrum. The first term, $ \widehat{h}(q,t=0)e^{\sigma(q) t}$, aligns perfectly with Eq.~\eqref{eq:lsa_linear_solution}, representing the spectrum of a thin film dewetting on a homogeneous substrate with contact angle $\theta_0$. The second term is proportional to the spectrum of the local wettability, $\widehat{\cos \theta}$. If the substrate has uniform wettability, this term vanishes. This is because if $\cos \theta$ is homogeneous, the amplitude of $\widehat{\cos \theta}$ is non-vanishing only for the zeroth mode, which does not affect any time evolution.
Otherwise, we see another contribution $\propto \widehat{\cos\theta}$ super-positioned on the homogeneous case spectrum. Because the initial film height $h_0$ is larger than the precursor length $h^*$ it holds by Eq.~\eqref{eq:wetting_potential} $\varphi'(h_0)<0$. Therefore, we obtain that the prefactor in front of $\widehat{\cos\theta}(q)$ is always positive, and we thus have, as expected, further excitement of all wavemodes $q$ whenever $\widehat{\cos\theta}(q)\neq 0$.

This maximum of $\sigma(q)$, known as the fastest-growing mode of the TFE on a plain substrate, is given by
\begin{align}
\label{eq:q_max}
q_{\max,plain}=\sqrt{\frac{1}{2}\varphi''(h_0)(1-(\cos\theta)_0)}.
\end{align}
The cutoff wavenumber of a thin film dewetting on a plain substrate, i.e., the largest wavenumber $q$ such that $\sigma(q)=0$, is given by
\begin{align}
    q_{0,plain}=\sqrt{2}q_{\max,plain}= \sqrt{\varphi''(h_0)(1-\cos\theta_0)}.
\end{align}
For all $q>q_{0,plain}$ it holds $\sigma(q)<0$. 

For every wavenumber $q$ such that $\widehat{\cos\theta}(q)>0$, it holds that the closer $q$ is to the intrinsic modes of the film, particularly the maximum $q_{\max,plain}$ of $\sigma(q)$, the more the wavemode of wavenumber $q$ is excited in the spectrum $\widehat{h}$. For large $q>q_{0,plain}$, one obtains for the plain dispersion relation $\sigma(q)<0$. $\sigma(q)$ is decaying rapidly as $-q^4$ for large $q$. Thus, the first term of Eq.~\eqref{eq:solution_linear_patterned_0} decays with time $t$ for $q>q_{0,plain}$. The larger $q$, the faster this decays over time. Nevertheless, for $\widehat{\cos\theta}(q)>0$, that wavenumber is still excited beyond the plain spectrum of a thin film dewetting on a chemically homogeneous substrate, but to a lower extent as for values of $q$ close to $q_{\max,plain}$. Accordingly, the modes of chemical patterning with wave numbers larger than the excited wavenumbers of the homogeneous-case thin-film are hard to detect in the dewetting film spectrum. Thus, defects of wavenumbers $q$, much larger than the cutoff wavenumber, $ q_{0,plain}$,  are not easily found in the spectrum, while wavenumbers $q\leq q_0$ can be frequently measured. The cutoff wavenumber $q_{0,plain}$ may vary depending on the liquid properties and especially on the height of the initial film. For an aqueous solution, it may range from $0.1\cdot1/\mu \text{m}$ to $100\cdot1/\mu\text{m}$~\cite{vrij_possible_1966}, enabling us to detect defects of patterns in the substrate as small as $0.01 \cdot \mu \text{m}$.

\ti{
%That chemical substrate patterns which resonate with the natural wavenumber $q_{\max,plain}$ of the thin film are more excited also reflects on the results of sections~\ref{sec:parameter_study}. 
For the majority of the simulations presented in Figures~\ref{fig:patterned_smoothed_pattern} and \ref{fig:sin_step_params}, one or two droplets (orange and yellow color code) per wavelength are observed. Those are the cases where the wavelength of the chemical patterning resonates with the wavenumber of the fastest growing mode of the film.  For these cases, the rupture time is reduced significantly from the plain reference case in those situations. In cases where the wavelengths of the substrate patterning and the fastest growing mode of the liquid film do not resonate, the rupture behavior resembles the plain reference case: For small wavelengths of the chemical patterning $\lambda_\theta=\omega_\theta$, we observe droplets spanning multiple $\lambda_\theta$ and longer rupture times. For large pattern wavelengths $\omega_\theta\ll q_{\max,plain}$, on the other hand, we observe many droplets per pattern wavelength $\lambda_\theta$ and increased rupture times.}

In Fig.~\ref{fig:spec}, the linear solution Eq.~\eqref{eq:solution_linear_patterned_0} is compared to the spectra (Fourier transform of the height field $\widehat{h})$) of the numerical solution of thin liquid films dewetting on one-dimensional substrate patterns and the linear solution on a "plain" substrate Eq.~\eqref{eq:lsa_linear_solution}. We show an early time step of the simulation. For different types of patterning, one can see that the linear theory and the actual full numerical simulation are in good agreement. Furthermore, the structure of the spectrum as the superposition of the "plain" spectrum and a multiple of the spectrum of the local wettability $\widehat{\cos\theta}$ is well demonstrated. In Fig.~\ref{fig:spec}, four different patterns were used. In panel (a), we show the spectrum of a thin liquid film dewetting on a chemically homogeneous "plain" substrate. In this case, the patterned LSA (Eq.~\eqref{eq:solution_linear_patterned_0}) and the standard LSA (Eq.~\eqref{eq:lsa_linear_solution}) coincide. \ti{Additionally, this serves as a validation case. It is well established that the LSA of the plain TFE is an accurate description of the spectrum of an early-stage dewetting liquid film by experiment~\cite{fetzer_thermal_2007} and molecular dynamics simulations~\cite{zhang_linear_2024, zhang_molecular_2019}, thus reproducing the expected spectrum validates our method.} Panel (b) shows the spectrum for the dewetting of a thin liquid film on a chemical pattern of the substrate that has been chosen such that $\widehat{\cos \theta}$ exhibits only a single excited wavenumber $q=1/\lambda_\theta$, where $\lambda_\theta$ is the wavelength of the used pattern. Remark that we are, consistently with our definition of the Fourier transformation Eq.~\eqref{eq:fourier_transform}, using the wavenumber $q=1/\lambda$ and not the angular wavenumber $2\pi/\lambda$. As a consequence, the spectrum obtained by Fourier transforming data from a numerical simulation of the situation, in agreement with Eq.~\eqref{eq:solution_linear_patterned_0}, is the spectrum of a thin liquid film dewetting on a plain substrate given by Eq.~\eqref{eq:lsa_linear_solution}, except for that one wavenumber of the pattern $1/\lambda_\theta$ that is additionally excited. In panel (c), we show the dewetting spectrum using a step-function pattern. Here, the most excited wavenumbers of $\widehat{\cos \theta}$ are all odd multiples of the pattern wavenumber $1/\lambda_\theta$. Our simulation data reproduce this. Finally, panel (d) shows the dewetting using a sine-wave pattern. The most excited wavenumbers here are all multiples of $1/\lambda_\theta$, and consequently, those are the wavenumbers we observe to be excited beyond the plain TFE solution. 

\section{Obtaining the substrate pattern from the thin film spectrum}
\label{sec:comparison}

\begin{figure*}
\centering
\includegraphics[width=\linewidth]{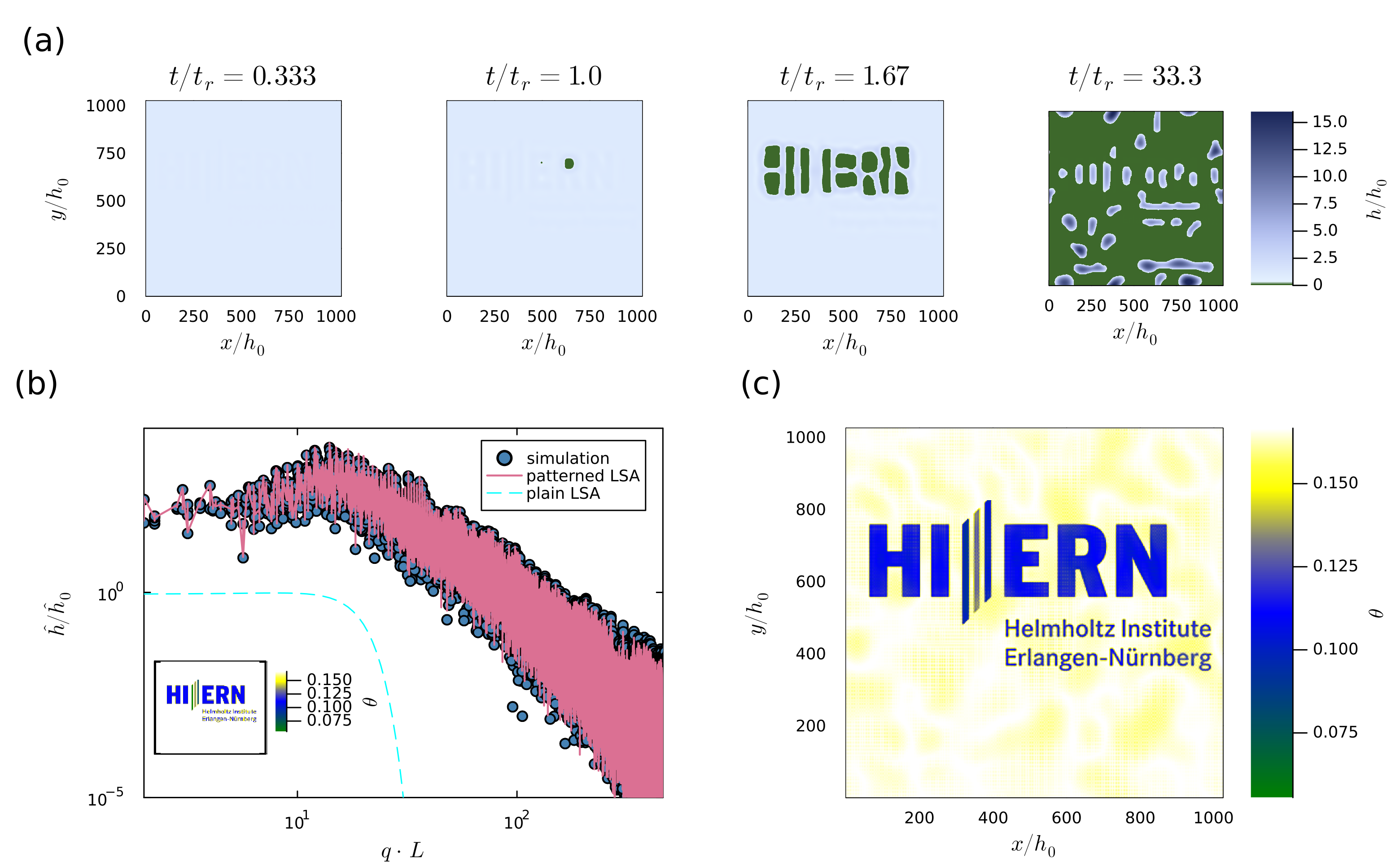}
\caption{Dewetting on two-dimensional patterned substrates.  (a) A simulation of a dewetting thin liquid film on a patterned substrate. The used chemical pattern is shown in the inset of panel (b). (b) The spectrum of a dewetting simulation of a thin liquid film on a two-dimensional chemical patterning (inset) shown as blue dots compared with the prediction of the LSA Eq.~\eqref{eq:lsa_linear_solution} (cyan line) and the prediction of the LSA including the pattern Eq.~\eqref{eq:paterned_fourier_0} (violet line). The data used corresponds to the first snapshot shown in Fig.~\ref{fig:patterned_3d}(a). (c) The chemical pattern obtained from the spectrum of the same simulation shown in Fig.~\ref{fig:patterned_3d}(a) by the reversed calculation Eq.~\eqref{eq:patterned_cos_theta}. The data used corresponds to the first snapshot shown in Fig.~\ref{fig:patterned_3d}(a) and  (b).}
\label{fig:patterned_3d}
\end{figure*}

Now that we have in-depth described the dewetting on thin liquid films on chemically patterned substrates and understood the initial film height evolution in the linear regime, we ask whether it is possible to invert the problem. We demonstrate that one can obtain information on the chemical patterning of the substrate while only having information about the liquid film on the top. 
Specifically, we show that the substrate pattern can be inferred from the spectrum of the liquid film. We then present the calculation necessary to do so alongside a numerical demonstration of their capabilities. We hope that this might inspire a new way of measuring local surface energies of substrates that extends beyond contact angle goniometry~\cite{kwok_contact_1999, vuckovac_uncertainties_2019} and scanning drop friction force microscopy (sDoFFI) \cite{hinduja_scanning_2022}. 

By measuring the spectrum of the film $\widehat{h}$ one can from Eq.~\eqref{eq:solution_linear_patterned_0} obtain the spectrum of the wettability pattern $\widehat{\cos \theta}$ by solving Eq.~\eqref{eq:solution_linear_patterned_0} for $\widehat{\cos \theta}$, yielding
\begin{align}
\label{eq:patterned_cos_theta}
	\widehat{\cos \theta}=-\frac{\widehat{h}(q,t) -\widehat{h}(q,t=0)e^{\sigma(q) t}}{(e^{\sigma(q)t}-1)q^2\gamma M(h_0) h_0}\sigma(q) \mu.
\end{align}
This step, of course, requires knowledge of all other physical parameters, such as the viscosity $\mu$, the surface tension $\gamma$, and the initial film height $h_0$. Furthermore, one needs good estimates for the initial perturbation $\widehat{h}(q,t=0) $ and the average wettability of the substrate $(\cos \theta)_0$, as well as the Hamaker constant, which determines the precursor height $h^*$ \cite{israelachvili_intermolecular_nodate}. Once one has the spectrum $\widehat{\cos \theta}$ an inverse Fourier transform ultimately leads to the substrate pattern $\theta=\cos^{-1}\left(\widecheck{\widehat{\cos \theta}}\right)$.  
\ti{Extending Eq.~\eqref{eq:patterned_cos_theta} to a two-dimensional height field is straight forward. We simply replace the wave-number $q$ with the wave-vector $\mathbf{q}$ yielding the Fourier transformations $\widehat{h}(\mathbf{q},t)$ and $\widehat{\cos \theta}(\mathbf{q})$. $\sigma(q)$ is calculated using the amplitude $q=\|q\|$. }
\ti{The described method provides the local apparent contact angle $\theta$, which is not only influenced by the solid-liquid surface energy but also by the local surface roughness through Wenzel's relation~\cite{de_gennes_capillarity_2004}.}
%Here, the hat $\widehat\quad$ notes the Fourier transform and the inverse hat $\widecheck\quad$ denotes the inverse Fourier transformation (see Eqs.~\eqref{eq:fourier_transform}, \eqref{eq:inverse_fouriere} and Ref.~\cite{evans2022partial}). 

To put Eq.~\eqref{eq:patterned_cos_theta} to the test, we perform a dewetting simulation on a two-dimensional pattern presented in Fig.~\ref{fig:patterned_3d}.
In Fig.~\ref{fig:patterned_3d}(a), snapshots from a simulation of the dewetting process on a patterned substrate are shown. The corresponding pattern is displayed in the inset of Fig.~\ref{fig:patterned_3d}(b) having regions with lower surface tension (blue letters and lowest in the green stripe) in the center of the pattern and higher contact angles (white regions) outside. In Fig.~\ref{fig:patterned_3d}(a), one can see holes nucleating, growing, and finally, droplets forming. At first glance, one may wonder why the film breaks first in the regions of lower surface tension (around the letters) and only then in the outer parts where the local contact angle is higher. This behavior is to be expected since for the step-function pattern in Fig.~\ref{fig:dewetting_patterned}(b), we already saw that valleys are first formed at the discontinuities of the pattern, i.e., in this case, at the pattern's outlines. Furthermore, the first hole to nucleate is located in the loop of the letter "R," where the local contact angle $\theta$ indeed is large. From the nucleation of the first holes, the dewetted contact lines then move outward. If one looks carefully, the substrate pattern is already visible in the simulation snapshots before rupture. 

In Fig.~\ref{fig:patterned_3d}(b), we compare the spectrum measured for the first snapshot of Fig.~\ref{fig:patterned_3d}(a) (i.e., at $t/t_r=1/3$) with the linear solution Eq.~\eqref{eq:paterned_fourier_0}. Again, the linear ansatz can explain the spectrum very well at this early time step. For this more random pattern, which includes all wave numbers in its spectrum $\widehat{\cos\theta}$ rather than just a few dominant ones like those in Fig.~\ref{fig:spec}, the spectrum of the dewetting liquid $\widehat{h}$ is less clearly related to the spectrum of dewetting on a plain, unpatterned substrate. Nevertheless, it stays true that the spectrum is the superposition of this plain spectrum $\widehat{h}(t=0)\exp(\sigma t)$ with a multiple of the spectrum of the wettability $\widehat{\cos\theta}$. All wavelengths are more excited than they would be without patterning, and rupture is significantly sped up. 
%In the supporting information, Fig.~S1, a plot of the same data with linear axes, including the mode $q=0$, is shown to demonstrate that for the mode $q=0$ one obtains $\hat{h}/\hat{h}_0=1$ for simulation, plain LSA, and patterned LSA. This is to be expected as the TFE \eqref{eq:TFE} is a conservation equation. 

Finally, we apply the inverted LSA Eq.~\eqref{eq:patterned_cos_theta} to the measured spectrum from Fig.~\ref{fig:patterned_3d}(b). To do so, no information about the chemical patterning is used. We only need the average $(\cos\theta)_0=\langle \cos \theta\rangle$. The result of this calculation is presented in Fig.~\ref{fig:patterned_3d}(c), recovering remarkable detail of the used chemical patterning. The shape of the pattern, even for the small letters, is perfectly recovered. Concerning the three stripes in the center of the chemical patterning, the absolute values deviate a bit towards the median value, i.e., the first stripe originally has a lower contact angle $\theta$ (green in the color scheme used) but is recovered from the spectrum with a too high contact angle (blue). At the same time, the second stripe with a high contact angle (yellow) is also shifted a bit towards the median contact angle (blue). Furthermore, the originally high homogeneous contact angle in the outer parts of the pattern is recovered a bit "stained." Nevertheless, the pattern can be recovered to a good extent and much sharper than visible from the dewetting simulation alone without applying Eq.~\eqref{eq:patterned_cos_theta}. 

In principle, it is feasible to measure the spectrum of film height in situ upon dewetting using an appropriate experimental setup. Such endeavors have been undertaken by Refs.~\cite{mu_dewetting_2000, meredith_combinatorial_2000, muller-buschbaum_dewetting_2002} using atomic force microscopy (AFM) and grazing incidence small-angle scattering (GISAS). In Ref.~\cite{becker_complex_2003}, very precise in situ measurements of thin film dewetting are performed by AFM. Also, X-ray reflection and refraction is an applicable method to measure the height spectrum of a liquid film with nanometric precision~\cite{butt_physics_2003}. Ref.~\cite{m_ller-buschbaum_dewetting_2003} provides a comprehensive review of how the film thickness and its spectrum may be measured by direct techniques such as AFM and with optical phase interference microscopy (OPIM) or reciprocal methods such as light, x-ray, or neutron scattering as well as GISAS, emphasizing the possibility of in situ measurements of the film height spectrum of a dewetting liquid film. To reproduce our theoretical results, calculating the surface energy pattern in terms of the local contact angle $\theta(x)$ from the film height spectrum as shown in Fig.~\ref{fig:patterned_3d}~(c), it is necessary to know not only the amplitude of the spectrum of the film height but also its phase. I.e., one requires the real and the imaginary part of the Fourier transform of the film height profile $\widehat{h}$. The AFM measurements of Ref.~\cite{becker_complex_2003} show that this is possible for viscous liquids, as the authors directly measure the film height profile. Upon Fourier transform, both the amplitude and phase of the spectrum are obtained. When using reciprocal scattering methods, one typically only obtains the amplitude but not the phase of the film height spectrum~\cite{m_ller-buschbaum_dewetting_2003}. From such measurements of only the amplitude of the film height spectrum, it is then possible to obtain the amplitude of the spectrum of the substrate patterning but not its phase. The inverse Fourier transform, yielding a real space map of the substrate contact angle $\theta(x)$, is thus not possible without the imaginary part of the spectrum. 

Nevertheless, applying our computational method to the experiments has some technical difficulties compared to the numerical simulation performed here. Wetting is a relatively fast process that can happen in a split second, complicating the measurements of the spectrum. This is typically cured by using very viscous liquids such as molten polystyrene or glycerol~\cite{fetzer_thermal_2007, fetzer_slippage_2007, becker_complex_2003, luo_ordered_2004}. With such materials, the dewetting process is prolonged to minutes, making it easier to observe in detail~\cite{becker_complex_2003}. Furthermore, not all input parameters may be known for an experiment. While the viscosity $\mu$ and surface tension $\gamma$ values are easily obtainable, the precursor height $h^*$ related to the Hamaker constant is difficult to assess, even though it can be calculated for certain simple liquids~\cite{israelachvili_intermolecular_nodate}. \ti{As shown in the supporting information (Fig. S3,S4,S5), a $\pm20\%$ error of the precursor height $h^*$ still leads to a recognizable image of the substrate patterning.} Also, the initial condition in an experimental setup will be less homogeneous and controlled. Thus, the initial film height $h_0$ and the initial amplitude of perturbation $\widehat{h}(q,t=0)$ will only be known with some uncertainty. \ti{We consider the measurements of Refs.~\cite{becker_complex_2003, fetzer_thermal_2007} most promising. The authors use $\sim 4 \text{nm}$ films of polystyrene spincasted on silicon wafers coated in silicon-oxide. Chemical patternings on the silicon wafer can possibly be created by etching and coating techniques.} How much of the results presented here can be achieved experimentally with different measurement and coating techniques is a fascinating question for further research. 

\section{Conclusion}
We studied the dewetting of ultra-thin liquid films on solid substrates that are physically flat but have a heterogeneous surface energy, investigating different patterns.
First, we went beyond the linear regime, studying the formation of droplet patterns after film rupture by a large numerical parameter study, transitioning from a step-function pattern to a sine-wave pattern and varying the wave number of the substrate pattern. In doing so, we observed several interesting transitions. For patterns with very small wavelengths, the emerging droplets span multiple periods of the tested pattern. Patterns with wavelengths around the spinodal wavelength of the same ultra-thin liquid film on a chemically homogeneous substrate show a transition between one and two droplets per pattern period after film rupture. The driving mechanism is the formation of holes in the liquid film, triggered by either spot or step-like defects. Finally, many droplets are observed per period of the test patterns for very large pattern wavelengths. When measuring the rupture time of the performed simulations, the regimes of different amounts of droplets after film rupture can be recognized, and the rupture time can be explained by the observed rupture dynamics. Thin films breaking on sine wave pattern were shown to result, independently of the pattern wavelength and amplitude,  in a single droplet per pattern period, except for the most extreme pattern-wavelengths considered here. For thin films breaking on step function patterns, on the other hand, either one or two drops emerge depending on pattern wavelength and amplitude.

We showed that the spectrum (by that, we mean the Fourier transformation of the film height profile) and rupture dynamics on a patterned substrate can be effectively characterized within the linear regime before rupture occurs using the Linear Stability Analysis (LSA). To do so, we also had to account for substrate wettability perturbations. It turns out that the spectrum of a film upon dewetting can be well approximated by the spectrum of a thin film dewetting on a plain substrate superpositioned with a multiple of the spectrum of the substrate pattern. That factor before the substrate pattern spectrum depends on the wavenumber and the liquid film's physical properties. 

By inverting our calculations, we proposed a method to obtain the spectrum of the underlying chemical pattern solely from information on the film height profile of a dewetting liquid film. By numerical simulations, we could show that the underlying pattern can be recovered in remarkable detail. This result holds potential for developing a novel method of surface energy measurements by measuring a thin liquid film dewetting on the substrate of interest. The feasibility of such a technique remains an exciting question for further research, which is underlined by the results of Refs.~\cite{hinduja_scanning_2022, choi_alternative_2003, becker_complex_2003}, which indicates that it might indeed be possible and beneficial to measure the surface energies of a substrate by the behavior of thin liquid films coated on top. 

\subsection*{Suporting Information}

For this article, there is supplementary material available. 

\subsection*{Data Availability}
The data that support the findings of this study are openly available in Zenodo at http://doi.org/10.5281/zenodo.XXXXXX.

\subsection*{Competing Interests}
The authors declare no competing interest.

\subsection*{Acknowledgments}
We acknowledge funding by the Deutsche Forschungsgemeinschaft (DFG, German Research Foundation)—Project-ID 431791331 (SFB 1452) and Project-ID 416229255 (SFB 1411).

% Bibliography
%\bibliography{resources}


\end{document}